\documentclass[aip,pop,floatfix]{revtex4-1}
\usepackage{fancyhdr}
\usepackage{upgreek}
\usepackage{setspace}
\usepackage[]{algorithm2e}
\usepackage{hyperref}
\usepackage{amsmath}
\usepackage{amsfonts}
\usepackage{amssymb}
\usepackage{graphicx}
\usepackage{subfigure}
\usepackage{verbatim}
\usepackage{bm}
\usepackage{rotating}
\usepackage{multirow}
\usepackage{booktabs}
\usepackage{graphicx} %use graph format  
\usepackage{epstopdf} 
\usepackage{dcolumn}% Align table columns on decimal point
\usepackage{bm}% bold math
\usepackage{subfigure}
\usepackage[switch,pagewise]{lineno}
\usepackage{color,soul}
\soulregister\ref7
\soulregister\cite7
\makeatletter

\newcommand{\Rmnum}[1]{\expandafter\@slowromancap\romannumeral #1@}
\makeatother
\usepackage{float}

\begin{document}

\title{Reconstruction of Poloidal Magnetic Fluxes on EAST based on Neural Networks with Measured Signals}
\setstretch{1} 
\setlength{\leftskip}{0em} \author{Feifei Long}
\affiliation{\setlength{\leftskip}{0em} \small Department of Plasma Physics and Fusion Engineering, University of Science and Technology of China, Hefei 230031, China}

\author{\setlength{\leftskip}{0em} Xiangze Xia}
\affiliation{\setlength{\leftskip}{0em} \small Department of Plasma Physics and Fusion Engineering, University of Science and Technology of China, Hefei 230031, China}

\author{\setlength{\leftskip}{0em} Jian Liu \thanks{Corresponding author: jliuphy@ustc.edu.cn}}
\email{Corresponding author Jian LIU: jliuphy@ustc.edu.cn}

\affiliation{\setlength{\leftskip}{0em} \small Department of Plasma Physics and Fusion Engineering, University of Science and Technology of China, Hefei 230031, China}
\affiliation{\setlength{\leftskip}{0em} \small Advanced Algorithm Joint Lab, National Supercomputing Center in Jinan, Jinan, Shandong 250014, China}

\author{\setlength{\leftskip}{0em} Zixi Liu}
\affiliation{\setlength{\leftskip}{0em} \small Department of Plasma Physics and Fusion Engineering, University of Science and Technology of China, Hefei 230031, China}

\author{\setlength{\leftskip}{0em}Xiaodong Wu}
\affiliation{\setlength{\leftskip}{0em} \small Institute of Plasma Physics, Chinese Academic Sciences, Hefei 230031, China}

\author{\setlength{\leftskip}{0em} Xiaohe Wu}
\affiliation{\setlength{\leftskip}{0em} \small Institute of Plasma Physics, Chinese Academic Sciences, Hefei 230031, China}

\author{\setlength{\leftskip}{0em} Chenguang Wan}
\affiliation{\setlength{\leftskip}{0em} \small Institute of Plasma Physics, Chinese Academic Sciences, Hefei 230031, China}

\author{\setlength{\leftskip}{0em} Xiang Gao}
\affiliation{\setlength{\leftskip}{0em} \small Institute of Plasma Physics, Chinese Academic Sciences, Hefei 230031, China}

\author{\setlength{\leftskip}{0em} Guoqiang Li}
\affiliation{\setlength{\leftskip}{0em} \small Institute of Plasma Physics, Chinese Academic Sciences, Hefei 230031, China}

\author{\setlength{\leftskip}{0em} Zhengping Luo}
\affiliation{\setlength{\leftskip}{0em} \small Institute of Plasma Physics, Chinese Academic Sciences, Hefei 230031, China}

\author{\setlength{\leftskip}{0em} Jinping Qian}
\affiliation{\setlength{\leftskip}{0em} \small Institute of Plasma Physics, Chinese Academic Sciences, Hefei 230031, China}

\author{\setlength{\leftskip}{0em}EAST Team}
\affiliation{\setlength{\leftskip}{0em} \small Institute of Plasma Physics, Chinese Academic Sciences, Hefei 230031, China}

\begin{abstract}
\textbf{Abstract}\\
 \setstretch{1} 
\small
\setlength{\leftskip}{0em}
The accurate construction of tokamak equilibria, which is critical for the effective control and optimization of plasma configurations, depends on the precise distribution of magnetic fields and magnetic fluxes. 
\textcolor{black}{Equilibrium fitting codes, such as EFIT relying on traditional equilibrium algorithms, require solving the Grad–Shafranov equation by iterations based on the least square method constrained with measured magnetic signals.
The iterative methods face numerous challenges and complexities in the pursuit of equilibrium optimization.
Furthermore, these methodologies heavily depend on the expertise and practical experience, demanding substantial resource allocation in personnel and time.}
This paper reconstructs magnetic equilibria for the EAST tokamak based on artificial neural networks through a supervised learning method. 
We use a fully connected neural network to replace the Grad-Shafranov equation and reconstruct the poloidal magnetic flux distribution by training the model based on EAST datasets. 
The training set, validation set, and testing set are partitioned randomly from the dataset of poloidal magnetic flux distributions of the EAST experiments in 2016 and 2017 years.
The accuracy of reconstructions is evaluated using a variety of indices, such as the mean squared error (MSE), peak signal-to-noise ratio (PSNR), and structural similarity index measure (SSIM), and similarity (S) with Fréchet distance. 
The feasibility of the neural network model is verified by comparing it to the offline EFIT results.
It is found that the neural network algorithm based on the supervised machine learning method can accurately predict the location of different closed magnetic flux surfaces at a high efficiency.
\textcolor{black}{The similarities of the predicted X-point position and last closed magnetic surface are both 98\%. The Pearson coherence of the predicted $q$ profiles is} \textcolor{black}{92\%.}
Compared with the target value, the model results show the potential of the neural network model for practical use in plasma modeling and real-time control of tokamak operations.
\end{abstract}
\keywords{\setlength{\leftskip}{0em} neutral network, poloidal magnetic flux, the last closed flux surface, X point, $q$ profile, EFIT, EAST}
\maketitle

\footnotemark{Code is available \textcolor{blue}{https://github.com/xz-xia/EAST-nns/}.} 

\setstretch{1.5} 
\section{Introduction}\label{sect1}
 
Poloidal flux structures in many tokamaks, such as EAST\cite{jinping2009equilibrium}, DIII-D\cite{lao2005mhd}, JET\cite{brix2008accuracy}, KSTAR\cite{jiang2021kinetic}, and NSTX\cite{gates2005plasma}, are usually reconstructed using EFIT, an MHD equilibrium code, which solves the Grad–Shafranov equation by iterations based on the least square method and measured experimental magnetic signals.
It takes at least 300 $\upmu$s for a GPU-based EFIT (P-EFIT) to complete one reconstruction iteration for the 65$\times$65 grid size on EAST\cite{huang2020gpu}.
However, future ITER needs to control the position and shape of the plasma on a higher time scale and possibly with a grid size exceeding 257×257 to maintain plasma stability. 
Therefore, the time cost for iterations of rt-EFIT is challenging to meet the real-time feedback control of plasma equilibrium parameters, including the poloidal flux, the toroidal current distribution, the minor radius, the position of the last closed flux, the magnetic axis, etc., over a duration of 1000 seconds.
Furthermore, the technical challenges of future reactors refer to not only the response time for long pulse operations but also different control requirements, such as the type-I edge localized mode (ELM) suppression using resonant magnetic perturbation coils\cite{park20183d}, divertor detachment scenarios\cite{reimold2015divertor,jaervinen2016comparison}.
Sophisticated control requirements rely on more precise and faster plasma equilibrium reconstruction tools, which can be applied in real-time operations. 

Artificial intelligence and machine learning have been rapidly developed and widely applied to fusion plasmas, such as automatic experimental data cleansing\cite{2019Time,2018An,2017Improvement,2017Preference}, structure-preserving simulation of plasma dynamical systems\cite{2022VPNets}, deep reinforcement learning magnetic control\cite{degrave2022magnetic}, plasma equilibrium control\cite{lao2022application}, global stability analysis\cite{piccione2020physics}, plasma disruption warning\cite{kates2019predicting}, automatic ELM-burst detection\textcolor{black}{~\cite{song2023development}}, L-H transition\cite{shin2018automatic}, the energy confinement scaling law\cite{chang2021constructing}, and the high-dimensional experiment database statistics\cite{zhu2020hybrid}.
It is well-known that the plasma equilibrium is vital for long-pulsed steady-state operations.
Without precise control, vertical instabilities and \textcolor{black}{disruptions\cite{ambrosino2005magnetic} } will bring a disastrous crash for a fusion reactor. 
The tokamak equilibrium involves multiple important physical properties, including the current, pressure, magnetic flux, position of the X point, and plasma shape.
Hence, precise equilibrium reconstruction is crucial for ensuring dependable real-time control and conducting authentic post-shot instability analysis.
\textcolor{black}{Over the past few years, KSTAR\cite{joung2019deep}, DIII-D\cite{lao2022application}, NSTX\cite{wai2022neural} develop several neural network models to extend plasma equilibrium reconstruction.
Two different loss functions were employed on KSTAR \cite{joung2019deep}, namely mean squared error (MSE) and a custom-defined loss function. 
By incorporating $\nabla$ $\psi$ into the loss function, two separate models with a spatial resolution of 22 $\times$ 13 matrices were trained. 
The first model solely provides accurate poloidal magnetic flux distribution, while the second model is capable of accurately predicting the current distribution. 
However, the low spatial resolution of 22 $\times$ 13 predicted by these two models is not suitable for input in simulations and experimental analyses.
Unlike KSTAR \cite{joung2019deep}, Lang Lao et al. \cite{lao2022application} in DIII-D trained three networks: one for predicting the $\psi$ distribution, another for predicting $\beta_{N}$, $\ell_i$, and $q_{95}$, and a third one for predicting LCFS. These three networks are integrated into a consolidated EFIT-MOR proxy model\cite{lao2022application}.
The biggest issue with the EFIT-MOR proxy model is that, for a single discharge, the predicted results between the three built-in models are likely to be inconsistent.
NSTX\cite{wai2022neural} develop two NNs models, i.e., E qnet and Pertnet, which is relevant to equilibrium and shape control modeling of fast prediction,
optimization, and visualization of plasma scenarios respectively.
Pertnet calculates the nonrigid plasma response while Eqnet is a reconstruction or forward free-boundary equilibrium solver.
In these studies, the $R^2$ (R-squared) coefficient is frequently employed as a benchmark for evaluating the performance of neural network (NN) models. 
Nevertheless, it is worth noting that relying solely on a high $R^2$ (R-squared) coefficient in regression values may not be the most suitable approach when assessing the congruence of distributions between the predicted outcomes from neural network models and the actual values.}

\textcolor{black}{EAST tokamak has utilized the back-propagation neural network to predict the minor radius, triangularity, elongation, upper X point, and lower X point of EAST tokamak using three different models as input parameters: 35 flux loops and 38 magnetic probes\cite{wang2016artificial}. 
However, the model did not directly predict the poloidal magnetic flux distribution, which may limit its applicability in the future.
Lu et al. trained a neural network model using the equilibrium data from EAST in 2022, with 35 flux loops, 34 magnetic probes, 14 poloidal field coil currents, and 1 plasma current as inputs, and the poloidal magnetic flux distribution as the output. The accuracy of the predicted results was evaluated using the normalized internal inductance. However, the use of the traditional mean squared error as the loss function and the normalized internal inductance as the evaluation metric, which is a volume integral, only allows for an overall assessment of the trend of the predicted results and cannot provide detailed differences \cite{lu2023fast}. 
In addition, Liu et al\cite{liu2022advanced} trained a model with Bayesian probability theory and neural networks to predict the current at different positions using 14 PF currents, 1 total plasma current, and 35 poloidal magnetic fluxes, and then calculated the poloidal magnetic flux distribution in reverse. 
Another work on EAST tokamak performs a 1D Swin-Transformer architecture and has also been used to predict the position of the last closed magnetic surface at different times. By taking control parameters such as 1 reference of plasma current, 1 in-vessel coil no.1 current, 12 poloidal field coils voltage, 12 nominal current of poloidal field coils, and 31 shape references as inputs, and using the measurement signals of 38 magnetic probes data and 35 flux loops data as outputs, a model result is obtained through training. Ultimately, this information is fed into the EFIT code to determine the position of the last closed magnetic surface.\cite{wan2023machine}  } 

In this work, we engage the fully connected neural networks (FNNs) to reconstruct the plasma poloidal flux on the 129 $\times$ 129 grid, which is consistent with the GS equation and the measured magnetic signals, based on machine learning. 
\textcolor{black}{Compared with the KSTAR, NSTX, and DIII-D models\cite{joung2019deep,wai2022neural,lao2022application}, we utilize a customized loss function that significantly improves the prediction of poloidal magnetic flux distribution}. By differentially weighting the regions inside and outside the last closed magnetic surface, we achieve a more accurate poloidal magnetic flux distribution. Consequently, this indicates the potential to acquire a more precise safety factor distribution.
Unlike iterating the GS equation with the least square method in EFIT, the NN method need not pre-calculate the green function matrix.
The Green functions compute magnetic fluxes on discrete grids in the $(R,Z)$ coordinate constrained with edge magnetic signal values.
Tokamaks with different architectures and parameters have different Green function lists.
Benchmarking EFIT inputs, the inputs of NNs in this work consist of 1 plasma current signal, 38 magnetic probe signals, and 35 flux loop signals.
The output is set to be a 129$\times$129 matrix, which provides the distribution of magnetic flux $\psi$ on grids.
The position of the X point and LCFS are then obtained.
The training set is generated using the off-line EFIT and EAST signals.
\textcolor{black}{The 129x129 grid is divided into two regions: one within the last closed magnetic surface, referred to as the "core region", and one outside the last closed magnetic surface, known as the "edge region." }
Since magnetic flux outside even beyond the vacuum vessel is of less significance, we aimed for greater accuracy in the magnetic flux distribution within the last closed magnetic surface. Therefore, a higher weight loss function is assigned to this area to enhance the network accuracy in this region.
In the NN model, we design a loss function with higher weight in the "core region" to improve the accuracy of NN-generated poloidal magnetic flux values.
It is worth noting that the training target is the numerical solution of the GS equation calculated by the off-line EFIT code, hence the prediction accuracy and other properties are also analyzed according to the off-line EFIT solutions.
Due to the static plasma equilibrium calculated by EFIT, there is no significant causal relationship between the previous moment and the following moment.  
Thus, whether shuffling the plasma equilibria data from various shots or shuffling different shots after bundling equilibrium data from multiple time slices within a single shot, neither method results in information leakage.
It should be noted that the neural network could predict better than real-time EFIT algorithms, but cannot exceed the baseline of the off-line EFIT on accuracy based on this dataset.
Once utilizing other diagnostic data as training features or constraints, machine learning methods will provide results beyond the scope of the GS equation and EFIT, since new physical or engineering information may be involved.

\setlength{\parskip}{0.5\baselineskip}
Because traditional iterations in the real-time EFIT reconstruction consume considerable computational resources, its instantaneity and accuracy should be further promoted.
On the contrary, machine learning methods based on NNs consume computational resources and data during offline training but perform swiftly at the same accuracy level in real-time applications.
In this paper, we generate distributions of poloidal fluxes on EAST utilizing fully connected neural networks.
We obtain three networks, named NN2016, NN2017, and NN20162017, respectively, which are trained using datasets from EAST discharges in different years.
All of them behave well on the testing set and have relatively good generalization capability.
Locations of the X-point and the LCFS can also be predicted according to the poloidal flux distribution output of neural networks.
Our model has been made publicly available on GitHub for researchers who are interested to explore and utilize it.
The rest part of the paper is organized as follows. 
Section~\ref{sect2} elaborates on the preparation of the dataset and the detailed reconstruction process using NNs. 
The performance and properties of the neural networks are analyzed in Sec.~\ref{sect3}. 
Finally, the discussion and plan are summarized in Sec.~\ref{sect4}.

\section{Neural network model}\label{sect2}
\subsection{Dataset}\label{sect2.1}
The EAST tokamak underwent significant upgrades and transformations involving changes to its first wall materials and divertor structures in the years 2014-2015. These modifications resulted in shifts in the locations of magnetic diagnoses. To account for the potential impact of changing diagnostic locations, we have chosen to focus our analysis on datasets collected after the substantial upgrades. Importantly, it is noteworthy that the positions of magnetic diagnoses within the facility remained unchanged during the subsequent years of 2016 and 2017. Therefore, our dataset exclusively includes information collected during these stable periods, allowing us to avoid potential complications arising from variations in diagnostic locations. We have utilized valid data specifically from these stable periods, obtained from the EAST facility.
\textcolor{black}{
Following chronological sequencing, we select discharges with plasma pulse duration falling within the range of 8 to 12 seconds, which results in the years 2016 and 2017 having 6298 shots and 2070 shots in the EAST tokamak, respectively.  
Therefore, following a ratio of 1:35  in the year 2016 and 1:15 in the year 2017, 182 shots and 138 shots are being filtered correspondingly from the respective totals of 6298 and 2070 shots. 
It should be noted that due to the engineering maintenance in the year 2017, the total number of plasma discharges in the year 2017 was fewer than that in the year 2016. 
The 320 plasma discharges are then shuffled and divided into training, validation, and test sets at a ratio of 0.8:0.1:0.1, which avoid some of the test and training dataset being gathered from the same discharges except different time slices.
The training process includes the time independent of the training set and test set, which excludes information leakage.
However, it is worth mentioning that the plasma equilibrium at a random time slice in the same discharge is independent since the EFIT code solves the static Grad-Shafranov equation without plasma transport. 
Due to the static nature of the equilibrium provided by EFIT in each shot, the states at two successive time points are independent variables. 
This is because the plasma is in a steady state ideal MHD equilibrium. 
Although the input and output parameters may exhibit similarities, there is no inherent correlation between them. }

Moreover, it should be noted that the reason for selecting the plasma pulse of  8 to 12 seconds stems from the fact that the discharge preparation phase usually sets a preset time of  10 to 12 seconds. 
If a plasma pulse can maintain a duration within 8 to 12 seconds after being set, it is considered that the plasma equilibrium during such discharges is relatively stable, indicating high-quality data. 
As each discharge spans 8-12 seconds, we segment the plasma equilibrium into different time slices, each with a 0.1-second interval, which results in each discharge having 80-120 plasma equilibrium, which leads to an imbalance in the number of balanced samples between the limiter configuration equilibrium and the divertor configuration equilibrium. 
In addressing the dataset's imbalance between the phases of the current ramp-up, flat-top, and descent, we introduce a sample-weight array during training.
This approach ensures a balanced mix of data, preventing an overrepresentation of plasma equilibrium in the flat-top phase of the plasma current and a shortage of data from the current rise and fall phases. 
This ensures that the network can learn effectively and produce more accurate outcomes, avoiding potential issues with suboptimal learning outcomes. 
\textcolor{black}{Before data cleaning, a raw database comprising 29,605 plasma equilibria is utilized from 320 shots.   
In the paper, a meticulous data preprocessing stage is performed incorporating human-assisted cleaning to ensure the quality and reliability of the dataset since the raw database contains some bad equilibrium samples, which are labeled with unclosed magnetic flux surface flux,  and magnetic reconnection distributions and so on. 
Bad examples may arise from diagnostic signal errors, the insufficient convergence of EFIT calculations, and other error sources.
Human-assisted cleaning played a crucial role in improving the dataset's quality, leading to a more precise neural network model. 
After data cleaning, 24955 equilibria in the 320 shots remain in the dataset, with 14189 samples originating from 2016 discharges and 10766 samples from 2017 discharges.
Among them, there are 2,697 samples in the current rise phase, 18,891 samples in the current plateau phase, and 3,367 samples in the current landing phase. }
 
\begin{figure}
  \centering
  \includegraphics[width=0.45\textwidth,height=0.5\textwidth]{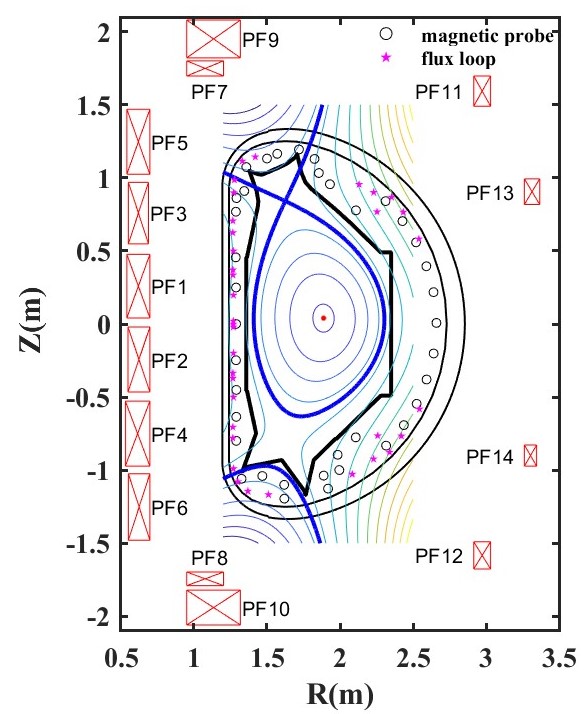}
   \caption{A poloidal cross-section of EAST equilibrium magnetic fluxes with the first wall. The bold blue curve represents the last closed flux surface in the upper single null configuration. The magnetic probes and flux loops locate outside of the first wall. }
   \label{fig1}
\end{figure}

In this paper, a total of 16641 poloidal magnetic flux values are distributed on 129$\times$129 spatial grids.
NN models are trained on the training set to adjust its parameters by minimizing the designated loss function. 
The validation set is utilized to evaluate the model performance during training and tweak hyper-parameters, which are established before model training, such as the learning rate and regularization strength. 
The validation set is useful for the model selection, to pick optimum hyper-parameters, and to detect overfitting or underfitting.
The testing set helps to testify performances of the trained model, such as the generalization capability and accuracy.
Furthermore, It should be noted that whether the time slices of the test set data and the training set data come from the same discharge or not, the change does not affect the reliability of the neural network prediction results because the individual time slices of each discharge are independent of each other and do not interact with each other.
The inputs of the model consist of one plasma current signal measured by the Rogowski coil, one toroidal field signal from the TF coil, 35 poloidal flux signals measured by flux loops, 38 magnetic probe signals from the measurement of local magnetic fields.
The total number of inputs is thus 75.
\textcolor{black}{The plasma current (Ip)  is the measurement of the external Rogowski to determine the total current in both the plasma and the vessel. Bt is the toroidal magnetic field from the toroidal field coils, which can be either parallel or antiparallel to the plasma current. 
EXPMP is a 1-D array containing an array of measured magnetic probe signals for each time point. 
Due to constant magnetic signals measured from 38 different locations  during the 2016 and 2017 years, the effect of the location in terms of the networks can be neglected by the input nodes.
COILS is a 1-D array of measured flux loops signals in v-sec/rad for each time point.}
Each example in the dataset is a 16716-dimensional vector (129$\times$129+75), consisting of 75 input values (attributes of each equilibrium) and 129 $\times$ 129 labeled output values ($\psi$ distribution on grids).
As shown in Fig.~\ref{fig1}, each black hollow circle indicates a magnetic probe, and each magenta pentagram denotes the position of a flux loop.  
The contour curves, based on 129 $\times$ 129 spatial grids, reveal the poloidal flux distribution, where the values of $\psi$ are derived from the off-line EFIT results.
The closed magnetic surfaces, which are depicted by concentric circles, are prediction targets of the NN model.

\begin{figure*}
  \centering
  \includegraphics[width=1\textwidth,height=0.5\textwidth]{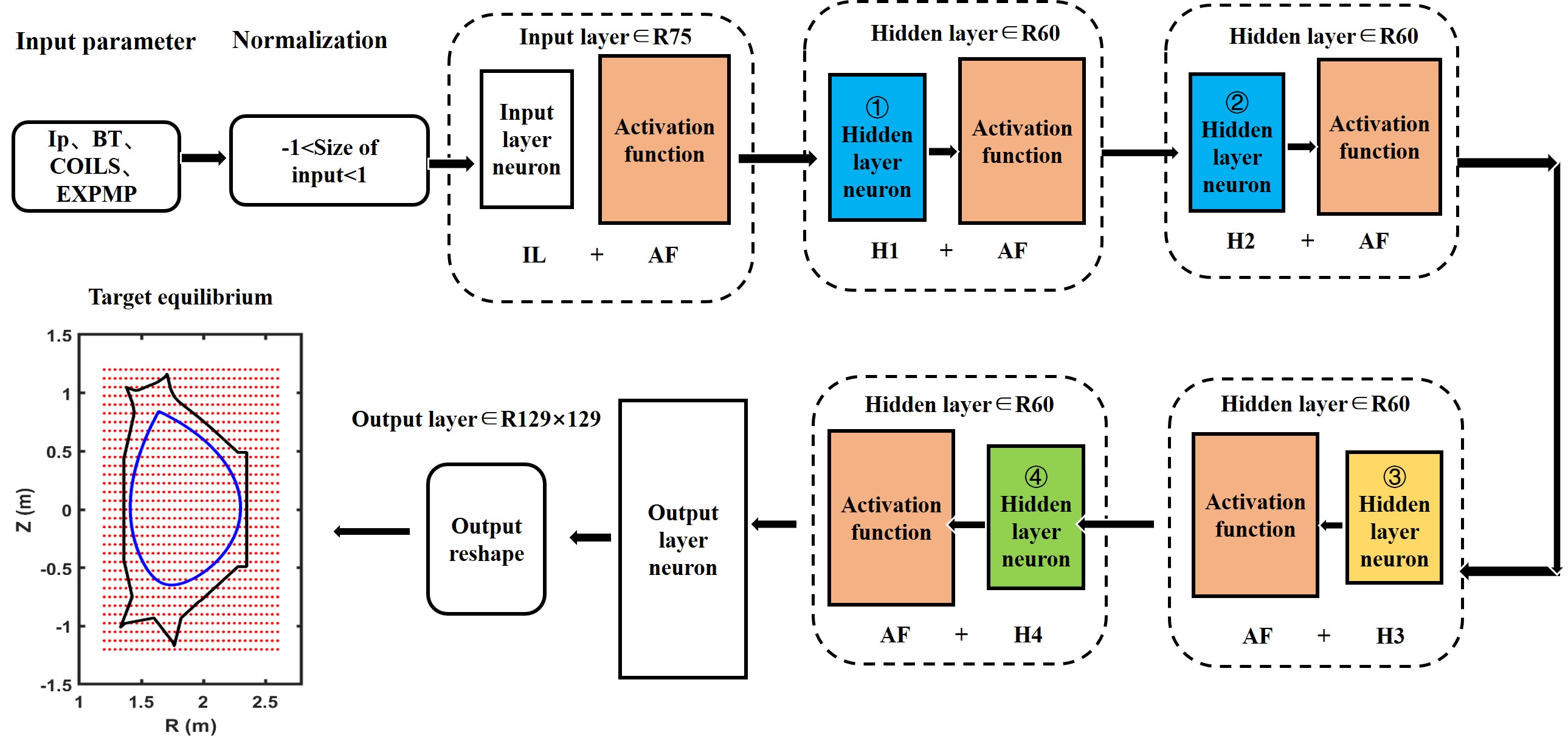}
   \caption{The neural network architecture and dataflow. The 75-dimensional input vector of the model consists of four groups, namely $I_{\rm p}$, $B_{\rm T}$, COILS, and EXPMP. The output is a poloidal flux distribution on 129 $\times$ 129 grids. The NN consists of one input layer, four hidden layers, and one output layer. }
   \label{fig2}
\end{figure*}

%==========================section 2.2%=======================

\subsection{Network model}\label{sect2.2}
A fully connected neural network (FNN) is engaged to generate values of $\psi$ on discrete spacial grids, according to the 75 input quantities.
As depicted by Fig.~\ref{fig2}, the network comprises an input layer with 75 nodes, four hidden layers each with 60 nodes, and an output layer with $129\times129=16641$ nodes.
Specific input quantities and the total number of examples used for model training are presented in Table~\ref{tab1}.
Consistent with the results of EFIT, outputs of the FNN should be a 129$\times$129 matrix.
Since inputs and outputs of FNNs are both vectors, the matrix needs to be flattened to a 16641-dimensional vector to fit the output of the FNN in practice.
It is found that an increase of hidden layers and hidden nodes can not significantly enhance the performances of this model.

\begin{table*}
   \caption{Information of inputs and examples.
    \label{tab1}}
\begin{center}
\begin{tabular}{|c|c|c|c|}
    \hline
Input Features & Explanation                                                                                              & Feature Size & Total number of examples          \\
    \hline
$I_p$         & the plasma current                                                                                     & 1      &       \\
\cline{1-3}
$B_T$        &the toroidal field at $R_{\rm maxis}$ & 1            &                        \multirow{2}{*}{29605 before data cleaning}                      \\
 \cline{1-3}
COILS      &\begin{tabular}{c}magnetic fluxes\\(Measured by flux loops)\end{tabular} & 35    &       \\
 \cline{1-3}
EXPMP    &\begin{tabular}{clc}magnetic fields\\(Measured by magnetic probes)\end{tabular} & 38       &       \multirow{3}{*}{24955 after data cleaning}                        \\ 
 \cline{1-3}
\multicolumn{2}{|c|}{Total number of input features}   & 75     &     \\ 
    \hline            
\end{tabular}
\end{center}
\end{table*}

A general FNN model can be explicitly forward calculated according to Eq.~\ref{eq3} as
\begin{equation}
y_{i}^{L+1}=\sigma \left[ \sum_{j=1}^{N_L}\left(w_{ij}^L y_j^L+b_{i}^{L+1} \right)\right]\ \ \ \ \ \ \ \ \ \ i\in\{1,2,3,...,N_{L+1}\} \ \ \ L\in\{0,1,...,M_o\}~,
\label{eq3}
\end{equation}
where $y_i^L$ denotes the value of the $i$-th node in the $L-$th layer, $N_L$ denotes the number of nodes in the $L$-th layer, $w_{ij}^L$ is the coefficient matrix between the $L$-th layer and $L_1-$th layer, $b_{i}^L$ is the bias for the $i$-th node in the $L$-th layer, and $\sigma(x)$ denotes the activation function.
In artificial neural networks, different activation functions are used to impart nonlinearity to mappings between adjacent layers.
Appropriate activation functions can reduce the vanishing gradient problem and facilitate the approximation capability and adaptability for a wide range of inputs.
For $L=0$, $y_i^0=x_i$ denotes the $i$-th input node.
For $L=M_o$, $y_i^{M_o}=\hat{y}_i$ denotes the $i$-th output node.
The number of hidden layers is $M_o-1$.
Then FNN can be represented by the nonlinear mapping $\hat{\mathbf{y}}(\mathbf{x})$. 

The FNN architecture in this work can then be written in Eq.~\ref{eq4} as
\begin{equation}
\setlength{\thickmuskip}{4mu}
\hat{\psi}_n=\sigma\left(b^5_n+\sum_{m=1}^{60} w^4_{nm} \sigma\left(b^4_m+\sum_{l=1}^{60} w^3_{ml} \sigma\left(b_l^3+\sum_{k=1}^{60} w_{lk}^2 \sigma\left(b_{k}^2+\sum_{j=1}^{60} w_{kj}^1\sigma\left(b_{j}^1+\sum_{i=1}^{75} w_{ji}^0 x_{i}\right)\right)\right)\right)\right)~,
\label{eq4}
\end{equation}
where \textcolor{black}{$n\in \{0,1,2,...,129\times 129\}$ is the index of output nodes, and the activation function is chosen as the $\tanh$ function}, i.e.,
\begin{equation}
\sigma(x)=\rm tanh(\emph x)=\frac{2}{1+e^{-2 \emph x}}-1~.
\label{eq5}
\end{equation}
%This FNN contains about one million free parameters to be determined, which is directly calculated as
%\begin{equation}
%N_{para}=75\times60+60\times60\times3+60\times16641+4\times60+16641=1,030,641~.
%\end{equation}
After training on the dataset, proper adjustments will be made and a precise FNN model will be established for the reconstruction of magnetic flux distributions.
In artificial neural networks, various activation functions are employed to introduce nonlinearity into the mappings between adjacent layers. These activation functions enable the network to capture and express nonlinear relationships within the data.
Appropriate activation functions can reduce the vanishing gradient problem and facilitate the approximation capability and adaptability for a wide range of inputs.

\subsection{Loss function}\label{sect2.3}
Loss functions determine the training goal of NNs and their capability to solve real problems.
We design two loss functions for the FNN model to solve the magnetic flux problem.
The first loss function $\varepsilon_{u}$ is a common mean squared error function governed by Eq.~\ref{eq1}.
The uniform loss function on a given set is explicitly defined to be
\begin{equation}
\varepsilon_{u}=\frac{1}{N}\sum_{n=1}^{N}{\Vert \hat\psi_n-\psi_n\Vert_2}^2~,
\label{eq1}
\end{equation}
where $n$ is the index of samples, $N$ is the total number of examples in the set, $\hat\psi_n$ denotes the output vector of the FNN for the $n$th input sample, $\psi_n$ denotes the labelled reference vector in the $n$th example. 
Both $\hat\psi_n$ and $\psi_n$ contain 16641 elements and can be expressed as a 129 by 129 matrix or their flattened form, i.e., a 16641-dimensional vector.
The symbol $\Vert\cdot\Vert_2$ denotes the 2-norm, which implies the Euclidean distance between $\hat\psi_n$ and $\psi_n$ in Eq.~\ref{eq1}.

Because the position of these input signals are measured ouside of the first wall shown as in the figure ~\ref{fig1}, the predicted $\psi$ value of the spatial range for the X and Y coordinate axes of the grid is 1.2 to 2.6 meters and -1.5 to 1.5 meters, respectively.
On the other hand, the precision of  $\psi$ within the last closed magnetic surface holds greater physical significance compared to its precision outside the last closed magnetic surface.
Hence,the edge elements of the matrix correspond to the first wall and the inner-outer vacuum region, where the distribution of magnetic fluxes is of no significance.
To improve the accuracy of $\psi$ values within the last closed flux surface, we design another loss function $\varepsilon _{core}$ as
\begin{equation}
  \varepsilon _{core}=\frac{1}{N} \left ( \sum_{n=1}^N \left[{\left\| W_L\cdot \left( \hat{\psi}_n-\psi _n \right) \cdot W_{R}^{T} \right\| _2}^2+ \Vert \hat\psi_n-\psi_n\Vert_2^2\right]\right) ,
  \label{eq2}
\end{equation}

where both $\hat\psi_n$ and $\psi_n$ take the form of 129 $\times$ 129 matrices, the left weight matrix $W_L$ and right weight matrix $W_R$ are both 129-dimensional column vectors defined through their components as
\begin{equation}
W_{Li}=\begin{cases}
	1, \,\,(m_r\leqslant i<n_r),\\
	0, \,\,\,\,else\\
\end{cases}~,\,\,
W_{Rj}=\begin{cases}
	1, \,\,(m_z\leqslant j<n_z),\\
	0, \,\,\,\,else\\
\end{cases}\,\,
~,
\label{eqomg}
\end{equation}
and $W_R^T$ denotes the transposition of $W_R$.
The vector index (i=$m_r$ to $n_r$ of R direction, j= $m_z$ to $n_z$ of Z direction) in Eq.\ref{eqomg} corresponds to the index value of the weight matrix,  separating the area inside the last closed magnetic surface from the area outside. 
This increases the weight of the poloidal magnetic flux inside the last closed magnetic surface, leading to a greater emphasis on predicting the position of the last closed flux surface. %$m_r$=46 $n_r$=88  $m_z$=34 $n_z$=100
The maximum value of the R-direction on the high magnetic field side of the plasma in the rectangular window ($m_r$, $n_r$, $m_z$, $n_z$) is slightly greater than that of the first wall, while the minimum value on the low magnetic field side is slightly smaller than that of the first wall. 
Additionally, the maximum value in the positive Z-direction of the rectangular window is slightly higher than the Z-value at the X-point, and the minimum value in the negative Z-direction is slightly lower than the minimum value at the X-point.

\begin{figure*}
  \centering
  \includegraphics[trim=0 20 20 20, clip,width=1\textwidth,height=0.45\textwidth]{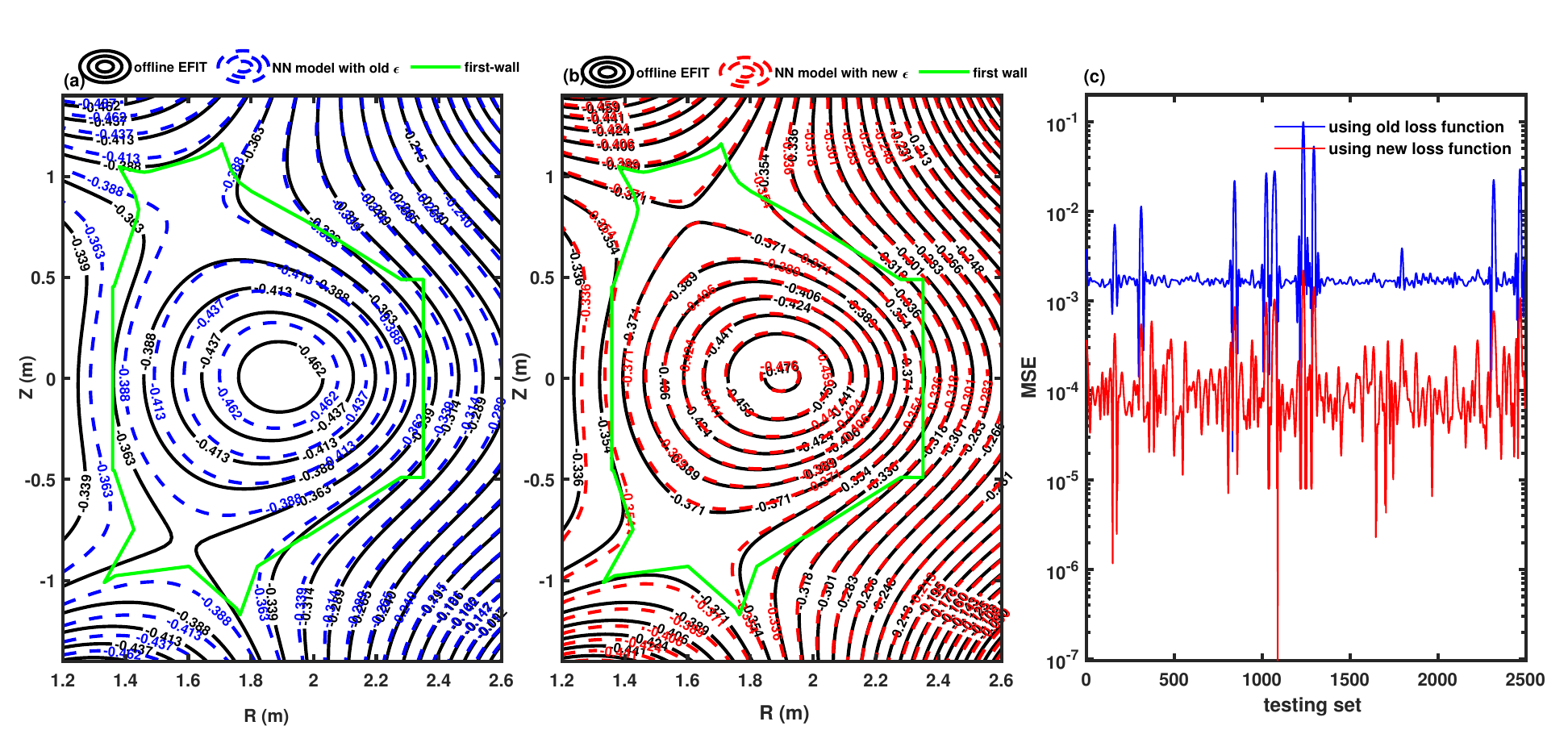}
   \caption{The FNN-generated poloidal flux distributions using different loss functions. Black curves are the reference fluxes calculated by the off-line EFIT. (a)Blue dashed curves are fluxes generated under the uniform loss function $\varepsilon_{u}$. (b)Red dashed curves are fluxes generated under the weighted loss function $\varepsilon_{core}$, which is more sensitive to the errors within the last closed flux surface.(c) The model's performance on testing set using  the old and new loss functions, where the MSE value is used to indicate the quality of predictive performance.}
   \label{fig4}
\end{figure*}

In Eq.~\ref{eq1}, errors of all elements of outputs are equally treated, while the weighted loss function in Eq.~\ref{eq2} is more sensitive to the errors of the elements within the last closed flux surface.
Figure~\ref{fig4} compares the results using the two different loss functions.
In Fig.\ref{fig4}(a), magnetic fluxes generated under the uniform loss function $\varepsilon_{u}$ deviate significantly from the reference fluxes.
For comparison, in Fig.\ref{fig4}(b), magnetic fluxes generated under the weighted loss function $\varepsilon_{core}$ are more accurate, especially within the last closed flux surface.
In Fig.\ref{fig4}(c), the model performance on the test set is used to compare the effects brought about by different loss functions.
When comparing the model's performance between weighted and unweighted loss functions on the test set using different loss functions, we found that the MSE values with the weighted loss function are significantly lower than those with the unweighted loss function, at least by an order of magnitude. This indicates that the weighted loss function has achieved better performance on the test set.
So the weighted loss function is employed in the FNN model.

\subsection{Network training}\label{sect2.4}
We train the FNN model by minimizing the weighted loss function $\varepsilon_{core}$ using the Adam optimization algorithm, which is a gradient descent method.
The Adam optimizer combines the Momentum and RMSprop algorithm and converges faster.
We use the mini-batch learning technique to perform the gradient descent process\textcolor{black}{, which is performed before training the network model.}
In this way,  the optimizer can escape from local minima and saddle points, and train the model faster.
The batch size is set to 64.
Small batches are partitioned from the 24955 examples after data cleaning, of which 19964 examples constitute the training set, 2495 constitute the validation set, and 2496 constitute the testing set.
The FNN model is carried out using the Tensorflow and Keras module in Python.

\begin{figure}
  \centering
  \includegraphics[width=0.5\textwidth,height=0.45\textwidth]{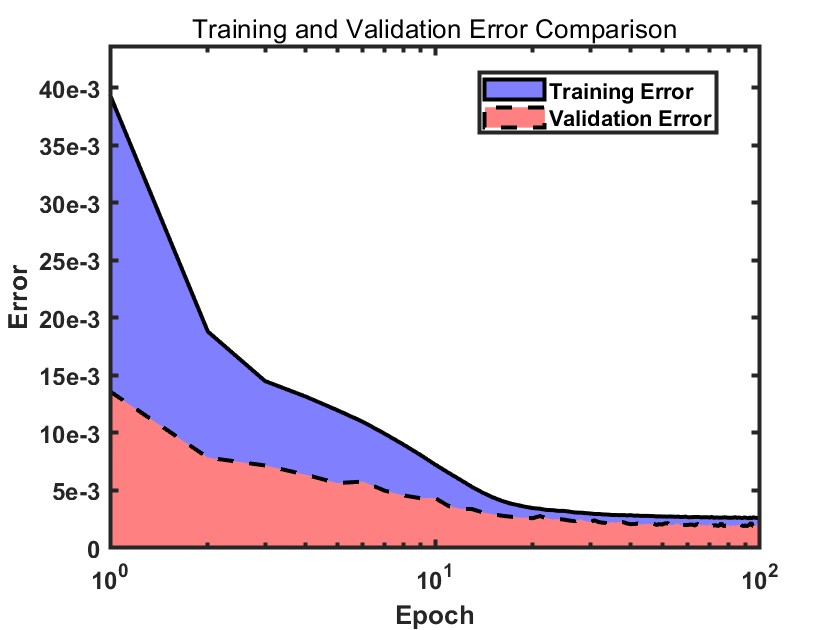}
   \caption{The learning curves include errors from the training set and validation set versus training epochs during network training.}
   \label{fig3}
\end{figure}

During pieces of training, we need to supervise the learning performance to estimate the convergence speed and the trend of error reduction.
Figure~\ref{fig3} plots the learning curves to help us better understand the training process of the model.
The solid and dashed curves represent the evolution of errors from the training set and the validation set, respectively.
With the increase of training epochs, both the training and validation errors keep decreasing.
After 100 epochs, both errors drop below $10^{-3}$, which indicates that the training and validation errors have already decreased to a rather low level and remained stable in the subsequent iterations.
It usually suggests that the model has been properly trained and not overfitting. 

\section{Performance and tests}\label{sect3}

After being properly trained, the performance of the FNN model are assessed in this section.
We utilize three key indices, i.e., mean squared error, peak signal-to-noise ratio, and structural similarity, to test the performances of the trained models.
MSE is widely used to measure the average difference between predictions and benchmarks and indicate the accuracy of the model. 
PSNR is commonly used in image processing to assess data quality and noise levels. 
SSIM is used to compare the similarity between two images, considering multiple factors such as brightness, contrast, and structure, providing a comprehensive evaluation of the similarity between matrices. 
%To evaluate the performance of the neural network, we can transform its output into a 129$\times$129 matrix and visually compare the predicted $\hat \psi$ with the $\psi$ calculated by the offline EFIT algorithm to generate an elliptical contour plot similar to Figure \ref{fig4}(b). 
Besides direct performance tests, we also examine physical results based on FNN reconstructions, such as the poloidal flux distributions, the position of the last closed flux surface, and the profile of the safety factor.
The tests on physical aspects provide intuitive verifications for the trained models and evaluate their utility values.

Furthermore, we train three FNN models on different training sets.
Based on the examples only from the 2016 EAST discharge campaigns, NN2016 is trained.
Based on the examples solely from 2017 EAST campaigns, NN2017 is trained.
And NN20162017 is trained on the 2016+2017 training set.
The differences among the three trained models reflect distinctions buried in data, including the position of diagnoses, ambient conditions, and experimental setups.
Characteristics of each campaign take its place in its own trained model.
We calculate MSE, PSNR, and SSIM for all three trained models to check their similarities and differences.

\subsection{Evaluation indicators}\label{sect3.1}
The MSE for a prediction of a well-trained FNN model is calculated according to 
\begin{equation}
\rm MSE=\frac{1}{129\times 129}\sum_{\mathit{i}=1}^{129}\sum_{\mathit{j}=1}^{129}[\hat\Psi(\mathit{i},\mathit{j})-\Psi(\mathit{i},\mathit{j})]^2,
  \label{eq6}
\end{equation}
where the $\hat\Psi(i,j)$ denotes the predicted $\psi$ value on grid $(i,j)$, while $\Psi(i,j)$ denotes the corresponding benchmark value, $i$ and $j$ here stand for grid indices. 
The formula calculates the mean squared differences between the predicted and target magnetic flux distribution point by point. 
The value of MSE ranges from 0 to positive infinity.
$MSE=0$ indicates that the prediction is exactly the same as the target distribution, while larger MSE implies poor prediction with greater errors. 

\begin{figure*}
  \centering
  \includegraphics[width=0.9\textwidth,height=0.4\textwidth]{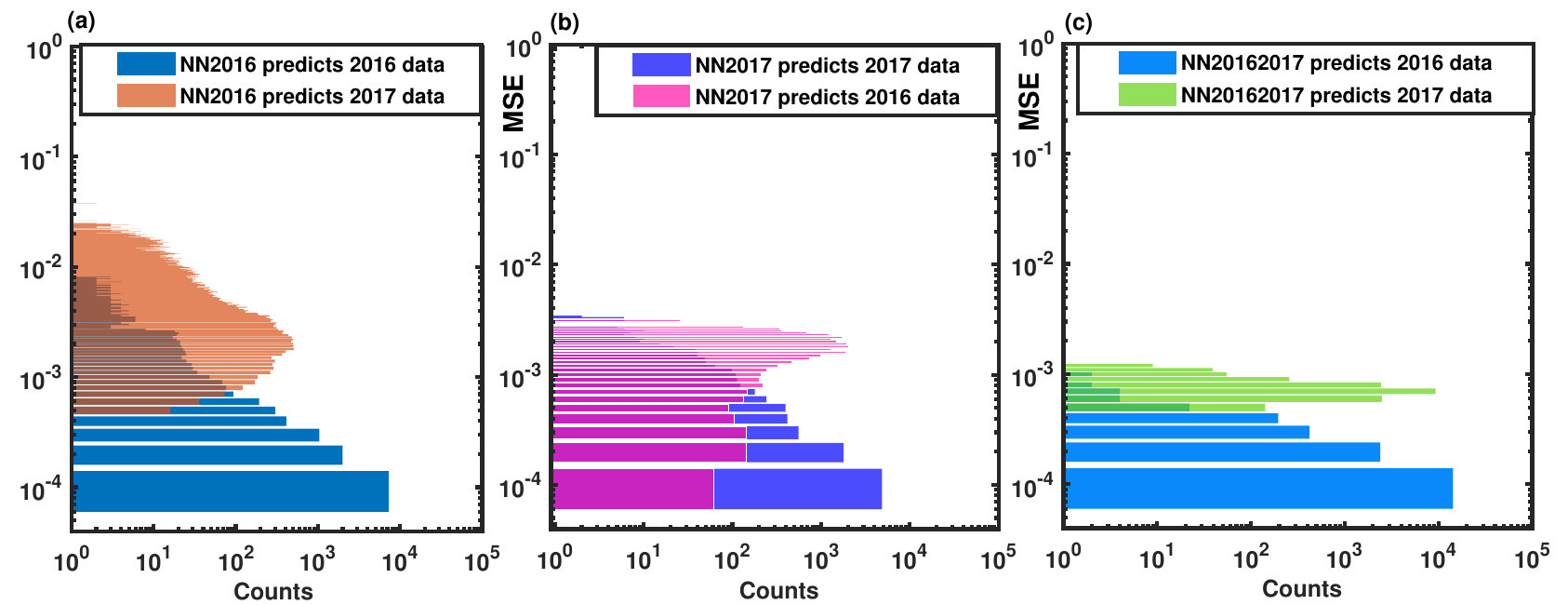}
   \caption{Mean squared error distributions on different datasets for different models, i.e., NN2016, NN2017, NN20162017, trained on datasets from different years. The Xlabel Counts represent the number of samples in the 2016 dataset, 2017 dataset, and 20162017 dataset, respectively. }
   \label{fig5}
\end{figure*}

Figure~\ref{fig5} plots histograms for counts of samples belonging to different error intervals from different models.
Testing errors on the 2016 dataset for NN2016 are relatively small. 
Similarly, testing errors on the 2017 dataset for NN2017 are also small. 
For comparison, testing errors on the 2017 dataset for NN2016 and testing errors on the 2016 dataset for NN2017 are relatively larger, which reflects different patterns of datasets from different years.
At the same time, NN20162017 performs best on the whole 2016+2017 dataset with all testing errors below the level of $10^{-3}$, which means that this model has the best generalization capability due to a larger and more diverse training set.
Overall, the performances of the three models on the 2017 dataset look poorer, though the scales of the 2016 and 2017 datasets are similar. 
It reflects that the data of 2017 are more complex, diverse, and representative.
This conclusion can also be inferred by observing that the MSEs of NN2017 on the 2016 dataset are much smaller than the MSEs of NN2016 on the 2017 dataset.
Generally speaking, the models trained on the 2017 dataset behave better, and the 2017 dataset provides a relatively better training set.

The PSNR of the predicted matrix is defined as
\begin{equation}
\rm PNSR=10\log_{10}{\frac{\psi_{\rm max}^2}{\rm MSE}}~,
\label{eq7}
\end{equation}
where $\psi_{\rm max}$ denotes the maximal element of the corresponding reference $\psi$ matrix, and $\mathrm{MSE}$ is the mean squared error provided by Eq.~\ref{eq6}.
The value of PSNR ranges from 0 to positive infinity, and its unit is decibel (dB).
It indicates the noise level of a signal and hence the similarity of a predicted matrix to the reference one.
Larger PSNR means better performance since the reciprocal of PSNR measures mean relative errors.
Typically, when the PSNR value exceeds 30dB, the difference between the predicted and target matrix is hard to perceive. 

The SSIM is calculated according to
\begin{equation}
\rm SSIM=\frac{(2\mu_{\hat \psi}\mu_{\psi}+C_1)(2\sigma_{\hat\psi\psi}+C_2)}{(\mu_{\hat \psi}^2+\mu_{\psi}^2+C_1)(\sigma_{\hat \psi}^2+\sigma_{\psi}^2+C_2)}~,
\label{eq8}
\end{equation}
where $C_1=k_1 L$ and $C_2=k_2 L$ are two variables to stabilize the division with weak denominator, $L$ is a constant which denotes the dynamic range of the reference matrix, $k_1=0.01$ and $k_2=0.03$ by default, $\mu_{\hat \psi}$ denotes the element average of the predicted matrix $\hat \psi$, $\mu_{\psi}$ denotes the element average of the reference matrix $\psi$, $\sigma_{\hat \psi}^2$ is the element variance of $\hat \psi$, $\sigma_{\psi}^2$ is the element variance of $\psi$, and $\sigma_{\hat\psi\psi}$ denotes the covariance between $\hat \psi$ and $\psi$.
The value of SSIM ranges from -1 to 1.
When the SSIM approaches 1, the two matrices become similar and the model performance is good.

\begin{figure*}
  \centering
  \includegraphics[width=1\textwidth,height=0.6\textwidth]{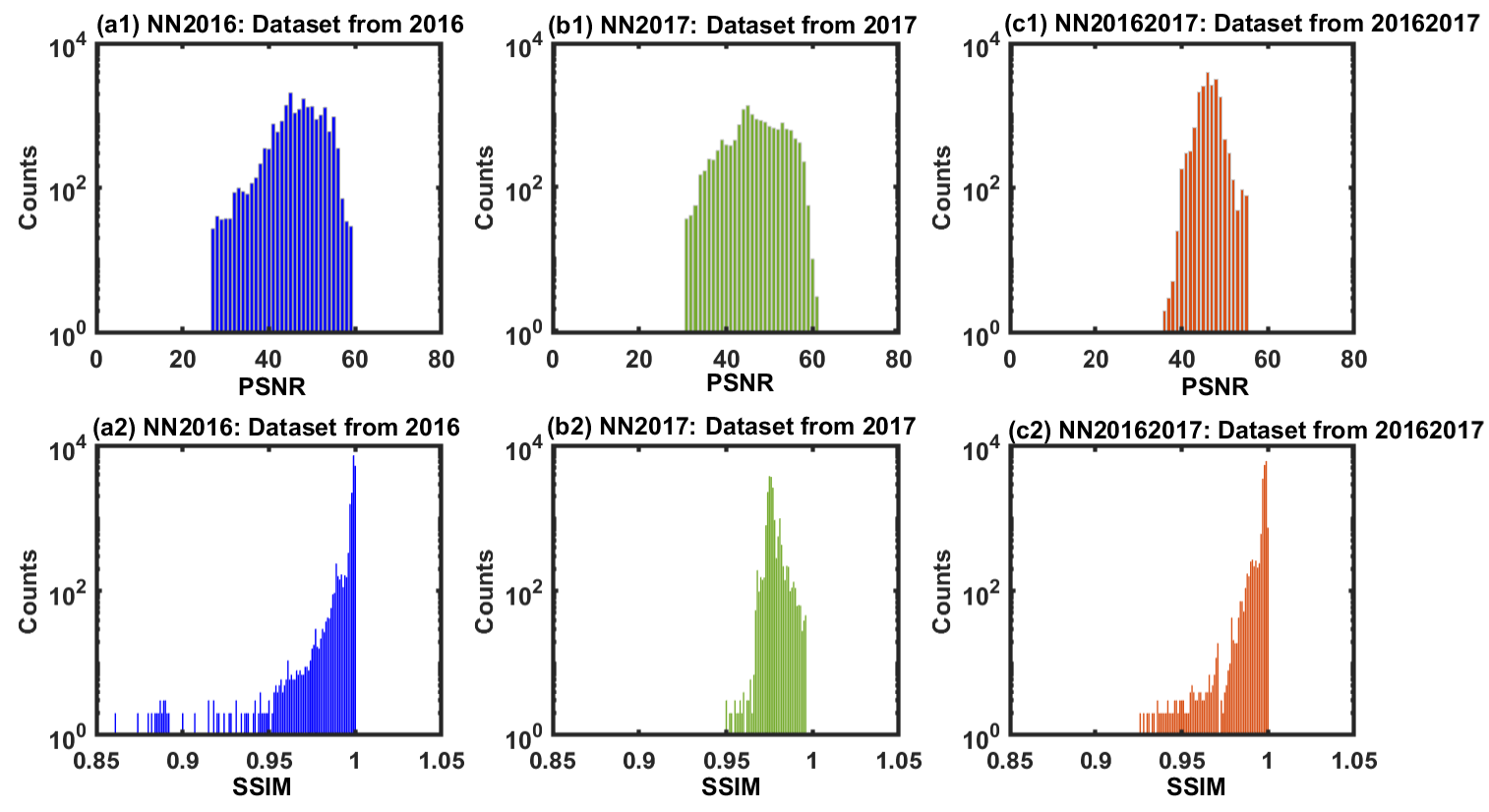}
   \caption{PSNR and SSIM distributions of NN2016, NN2017, and NN20162017 on different  datasets. The blue bars represent the performance of NN2016 on the 2016 dataset, the green bars represent the performance of NN2017 on the 2017 dataset, and the orange bars represent the performance of NN20162017 on the 2016 plus 2017 dataset. }
   \label{fig6}
\end{figure*}

PSNRs and SIMMs for different trained models are compared in Fig.~\ref{fig6}.
The PSNR distribution of NN20162017 looks more concentrated, compared with NN2016 and NN2017.
 And the SSIM distribution of NN20162017 is overall closer to 1.
 All these results verify the good performance of NN20162017.
 The PSNR performance of NN2016 and NN2017 looks similar, but their SSIM distributions exhibit different properties.
 More samples in NN2016 have SSIM closer to 1, while the number of samples with small SSIM in NN2016 is also larger than in NN2017.
 Most samples in NN2017 have SSIM around 0.97.
 The shape of the SSIM distribution for NN2017 significantly differs from that for NN2016 and NN20162017.

\subsection{X-point position and last closed magnetic surface}\label{sect3.2}
\begin{figure*}
  \centering
  \includegraphics[trim=20 10 10 5, clip,width=1\textwidth,height=0.5\textwidth]{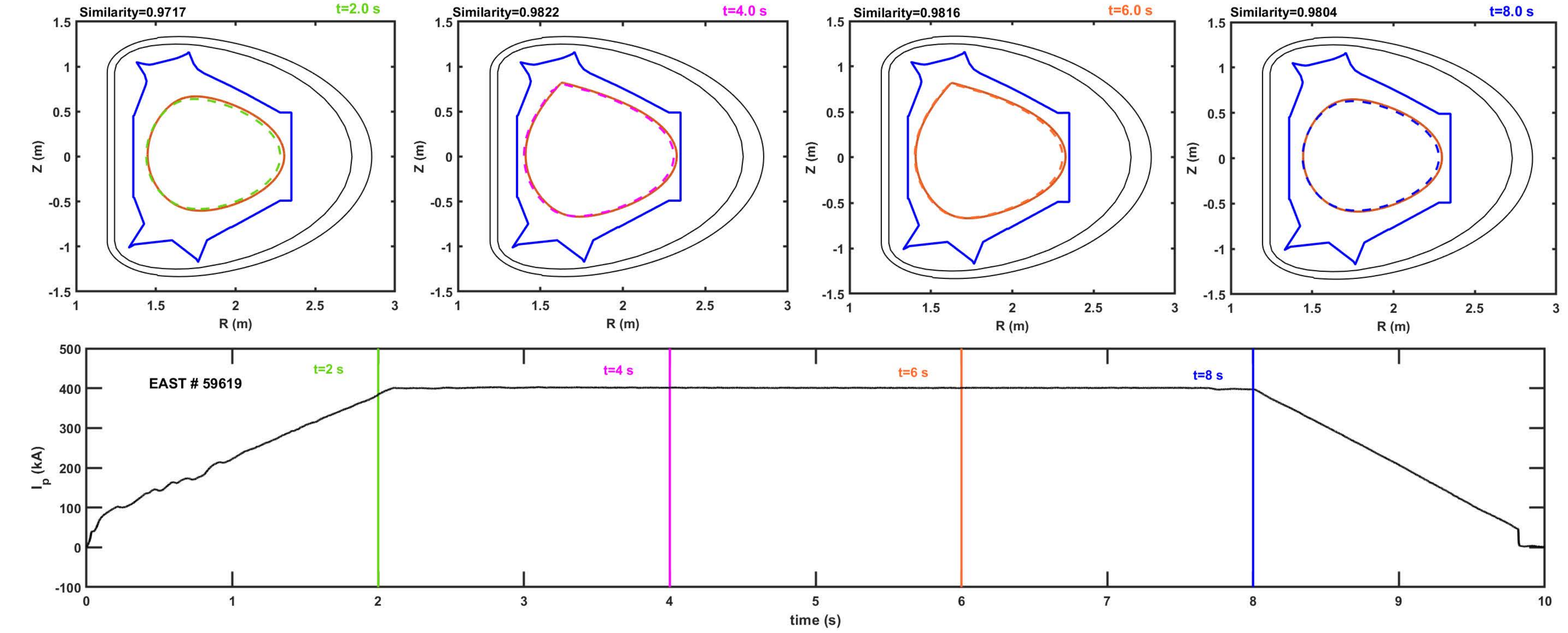}
   \caption{Discharge $\#59619$ in the test set on the comparison between offline EFIT LCFS reconstruction and the NN20162017 prediction. The LCFS was generated from off-line EFIT using inputs from our model. The solid blue lines are the experimental target LCFS, the dashed lines in  different color  are predicted LCFS. Note that the all plasma equilibria in the shot $\#59619$ used for testing are not included in the training set and are exclusively reserved for testing on the test set. }
   \label{fig7}
\end{figure*}
Besides accurate $\psi$ distribution matrices, we intend to obtain precise locations of the X-point and the last closed magnetic surface from the output of FNN models.
We choose the best-trained model among the three, i.e., NN20162017, to complete magnetic flux reconstructions.
To evaluate the accuracy of the predicted location of X-points, we compare the coordinates of X-points calculated from FNN outputs with corresponding reference values using their absolute errors
In the poloidal plane, the location of the X-point is written in the $R-Z$ components of a cylindrical coordinate system.
Figure~\ref{fig7} plots the FNN-predicted LCFSs at three different time slices in the $\#59619$ EAST discharge, compared with benchmark solutions.
\textcolor{black}{ The discharge $\#59619$ is solely used for the testing set and all of the time slices from the discharge are not used for the training and test data as well.}
Corresponding similarity values $S=1/(1+d_F)$ are calculated, where $d_F$ denotes the Fréchet distance.
The Fréchet distance between two sets of samples is defined by minimizing the maximum distance between any two corresponding points on the two curves. 
It can be observed that the LCFS estimation errors of NN20162017 for the $\#59619$ EAST discharge are rather small, with $S=0.9717$ at t=2s, $S=0.9822$ t=4s,  $S=0.9816$ at t=6s,and $S=0.9804$ at t=8s, respectively.

\begin{figure}
  \centering
  \includegraphics[width=0.5\textwidth,height=0.7\textwidth]{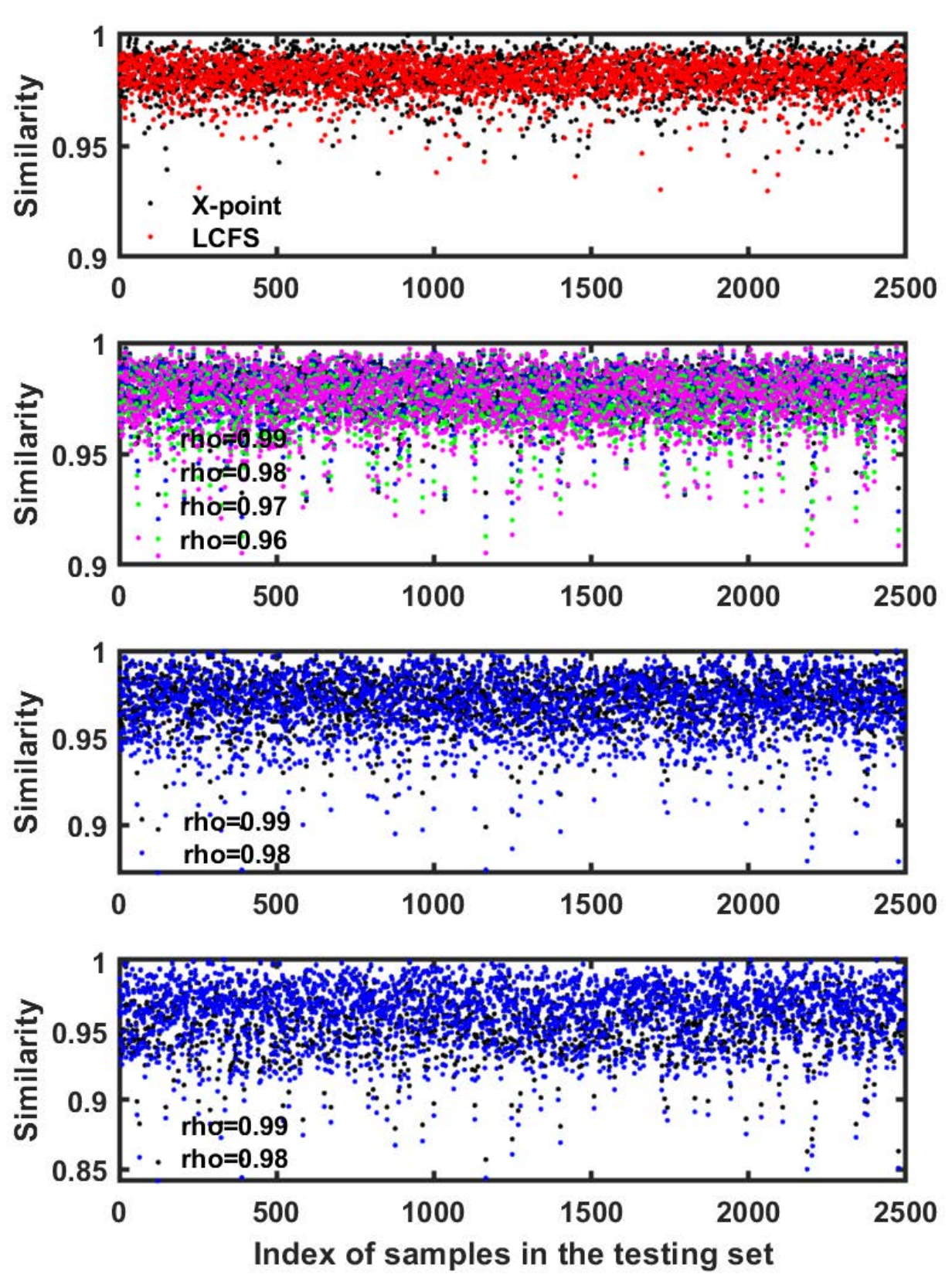}
   \caption{The distribution of similarity values for X-point locations, LCFSs, and core magnetic flux surfaces at different normalized radius $\rho$ positions between NN20162017 predictions and references in the testing set. }
   \label{fig8}
\end{figure}

We also compute the accuracies of X-point positions and LCFSs on the testing sets. 
The similarities of X-point locations and LCFSs between FNN outputs and corresponding benchmark values are plotted in Fig.~\ref{fig8}. 
The mean similarity of X-point locations as well as LCFSs is about 0.98, which is accurate enough for most practical applications.
Furthermore, similarities of core magnetic flux surfaces at the different normalized radius $\rho$ are also presented. 
The mean similarity of core flux surface predictions in the range of $\rho=0.90-0.99$ is still around 0.98, while it decreases slightly when $\rho=0.8$.
Physically, the core magnetic surfaces exhibit more deviations due to the lack of extra constraints within the last closed flux surface. 
These tests verify that the NN20162017 model exhibits good performances not only on the generation of poloidal flux distributions, but also on predictions of the X-point location, LCFS, and closed magnetic surfaces at different position.

\subsection{$q$ profile}\label{sect3.3}
The safety factor is a key factor in magnetized confinement fusion devices and has a vital impact on the stability and transport of plasma in tokamaks.
Higher safety factor results in better stability, where MHD instabilities can be suppressed more easily.
Therefore, the distribution of safety factors is crucial for the suppression of stabilities and control of tokamak plasma.
The safety factor  $q$ can be defined as
\begin{equation}
q=\frac{\mathrm{d}\phi}{2\pi \mathrm{d}\psi}~,
\end{equation}
where $\phi$ denotes the toroidal magnetic flux, and $\psi$ is the poloidal magnetic flux.
So the safety factor is the rate of change of the toroidal magnetic flux with the poloidal magnetic flux.
\begin{figure*}
  \centering
  \includegraphics[width=0.95\textwidth,height=0.8\textwidth]{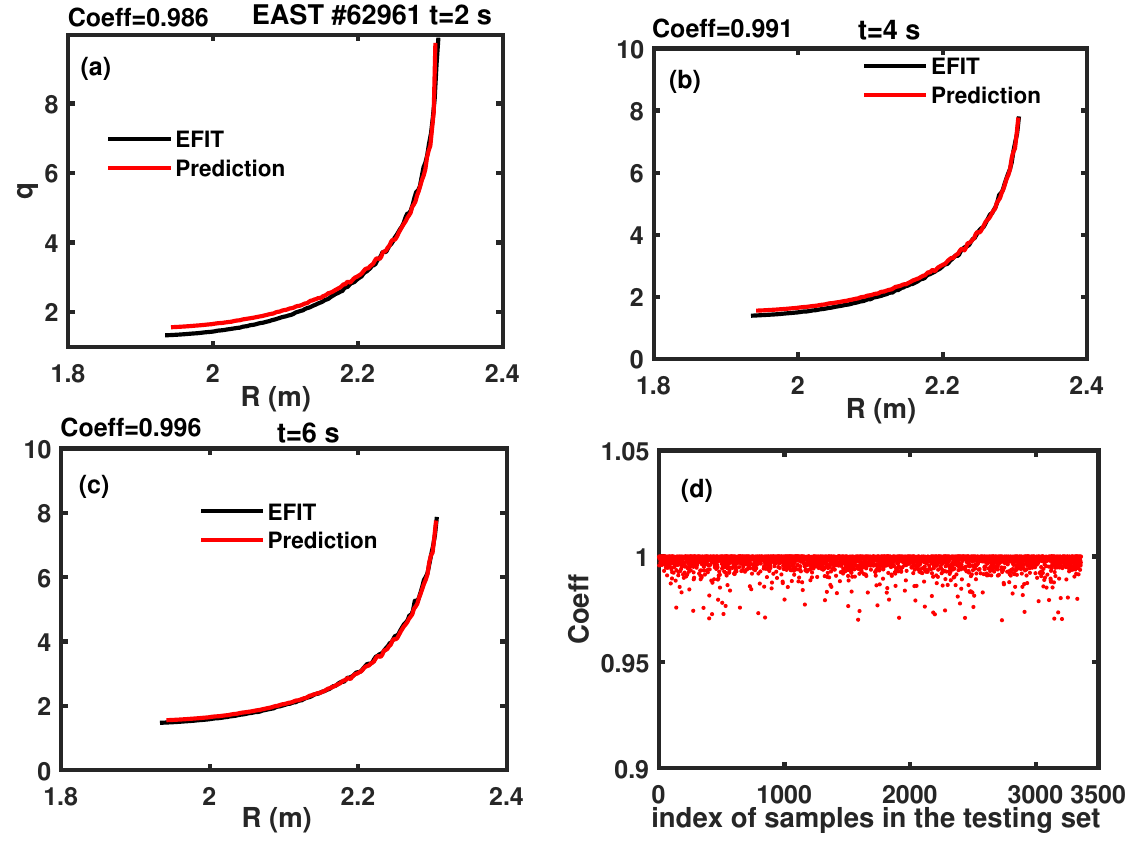}
   \caption{$q$ profile reconstruction on the EASU discharge $\#62921$ in the testing set. The $q$ in black lines are generated from off-line EFIT using inputs from our model. The red lines denote the results from the NN20162017 predicted poloidal magnetic flux. The $Coeff$ denotes the Pearson correlation coefficient. }
   \label{fig9}
\end{figure*}
Because the toroidal magnetic field is largely determined by the engineering design and is essentially constant, we treat it as an input variable. 
Therefore, the safety factor profile is related to the distribution of the poloidal magnetic flux.
In addition, as q is not a position in geometric space, it is not suitable to calculate the similarity using the Frechet distance. 
Therefore, for q distribution, we use the Pearson correlation coefficient as the evaluation criterion.
The formula for the Pearson correlation coefficient is as follows:
$$Coeff = \frac{cov(\psi^{\rm true},\psi^{\rm pred})}{\sigma_\psi^{\rm pred} \sigma_\psi^{\rm true}} = \frac{\sum_{i=1}^n (\psi^{\rm true}_i - \bar{\psi}^{\rm true})(\psi^{\rm pred}_i - \bar{\psi}^{\rm pred})}{\sqrt{\sum_{i=1}^n (\psi^{\rm true}_i - \bar{\psi}^{\rm true})^2} \sqrt{\sum_{i=1}^n (\psi^{\rm pred}_i - \bar{\psi}^{\rm pred})^2}}$$
In this formula, $Coeff$ represents the Pearson correlation coefficient between variables $\psi^{\rm true}$ and $\psi^{\rm pred}$, $cov(\psi^{\rm true},\psi^{\rm pred})$ represents the covariance between $\psi^{\rm true}$ and $\psi^{\rm pred}$, and $\sigma_\psi^{\rm true}$ and $\sigma_\psi^{\rm pred}$ represent the standard deviations of $\psi^{\rm true}$ and  $\psi^{\rm pred}$, respectively. $\bar{\psi}^{\rm true}$ and $\bar{\psi}^{\rm pred}$ represent the means of $\psi^{\rm true}$ and $\psi^{\rm pred}$, respectively.

Figure~\ref{fig9}(a)-(c) shows the reconstruction of $q$ profiles at t=2 s, 4 s, and 6 s in the $\#62921$ EAST discharge using NN20162017 compared with the reference values. 
The corresponding correlation coefficients are computed at the same time.
The results verify that the $q$ profiles predicted by the FNN model are in excellent agreement with the results from EFIT for three different times.
Figure~\ref{fig9}(d) plots the distribution of Pearson correlation coefficient of $q^{true}$ and $q^{pred}$ profiles on the testing sets. 
The mean correlation coefficient of model-predicted q profiles is about 0.9964 on the testing sets.

\section{Discussion and Summary}\label{sect4}
\begin{figure*}
  \centering
  \includegraphics[trim=10 10 10 0, clip, width=1.1\textwidth,height=0.5\textwidth]{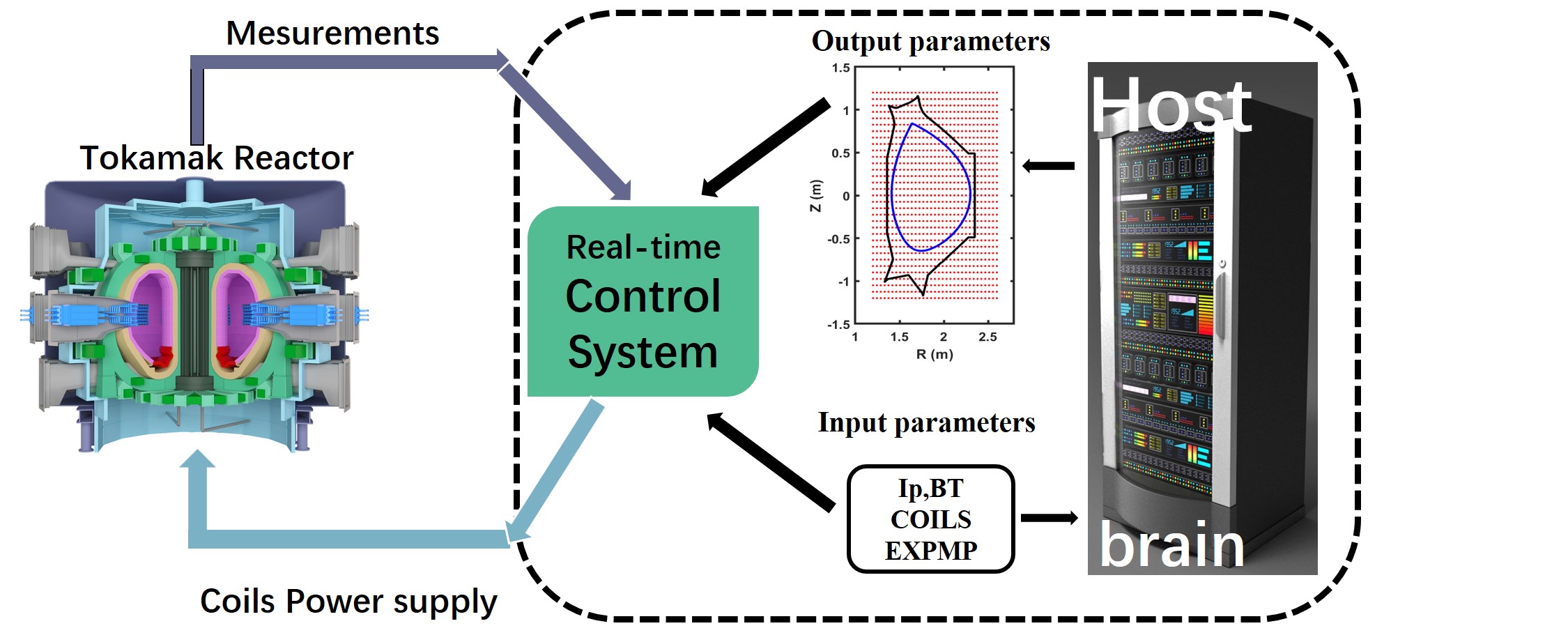}
   \caption{Diagram of the practical application of trained FNN models in real-time reconstructions, operations, and automatic controls of tokamak devices.}
   \label{fig10}
\end{figure*}
In this paper, we train three FNN models to reconstruct poloidal magnetic flux distributions according to EAST experimental data and test their performances on different testing sets.
A variety of indices, including MSE, PSNR, and SSIM, are utilized to analyze the models' behaviors.
The positions of X-points, LCFSs, core flux surfaces, and q-profiles are also given by the magnetic flux surfaces reconstructed from the FNN model.
These results verify that the trained FNN models can reconstruct the poloidal flux distributions, as well as related physical quantities, at a precise level, which is crucial for the design and optimization of tokamak operations. 
The time cost for the reconstruction of one flux distribution using the FNN model is about 75 $\mu s$ running on a normal laptop, which is competitive for the usage of real-time equilibrium reconstruction during operations and real-time controls, see Fig.~\ref{fig10}.
Compared with traditional software, the NN models have great cross-platform portability, because all different hardware architectures, such as GPU and many-core chips, have their own machine-learning framework.
The trained model can be directly deployed and run on different platforms, without rewriting massive codes with different computer languages.
Good support and optimization for neural networks enable us to promote the efficiency of NN models conveniently.
In future work, we will further improve the accuracy and efficiency of the FNN model and make efforts to integrate it into the real-time operation workflow.

\section*{Acknowledgements}
This research is supported by
the National Natural Science Foundation of China (Nos.12171466, 12175277, 11975271), the National Magnetic Confinement Fusion Energy Research and Development Program of China (Nos.2019YFE03090100, 2019YFE03060000, 2022YFE03050003), and Geo-Algorithmic Plasma Simulator (GAPS) Project.
%\section*{References}

%\bibliographystyle{aipnum4-1}
%\bibliographystyle{aipauth4-1}
\bibliography{refxxz}

\end{document}